\documentclass{metalcone}
\usepackage[super,comma,sort&compress]{natbib}
\usepackage[T1]{fontenc}
\usepackage[flushleft]{threeparttable}
\usepackage{hyperref}
\hypersetup{colorlinks,citecolor=blue}
\usepackage{mathptmx}
\usepackage{multirow}
\usepackage{amsmath}

\begin{document}
\setlength{\parskip}{1em}
\pagestyle{plain}

\justifying
\title{Disorder controlled sound speed and thermal conductivity of hybrid metalcone films}

\maketitle


\author{Md Shafkat Bin Hoque, Rachel A. Nye, Saman Zare, Stephanie Atkinson, Siyao Wang, Andrew H. Jones, John T. Gaskins, Gregory Parsons, and Patrick E. Hopkins}

\begin{affiliations}
\normalsize Md Shafkat Bin Hoque, Saman Zare\\
Department of Mechanical and Aerospace Engineering, University of Virginia, Charlottesville, Virginia 22904, USA\hfill

\normalsize Rachel A. Nye, Stephanie Atkinson, Siyao Wang, Gregory Parsons\\
Department of Chemical and Biomolecular Engineering, North Carolina State University, Raleigh, North Carolina 27606, United States\hfill

\normalsize Andrew H. Jones, John T. Gaskins\\
Laser thermal Inc., Charlottesville, Virginia 22902, USA\hfill

\normalsize Patrick E. Hopkins\\
Department of Mechanical and Aerospace Engineering, University of Virginia, Charlottesville, Virginia 22904, USA\\
Department of Materials Science and Engineering, University of Virginia, Charlottesville, Virginia 22904, USA\\
Department of Physics, University of Virginia, Charlottesville, Virginia 22904, USA\\
Email: phopkins@virginia.edu

\end{affiliations}


\dedication{}

\dedication{}


\newpage
\textbf{\Large Abstract}
\begin{abstract}

The multifaceted applications of polymers are often limited by their thermal conductivity. Therefore, understanding the mechanisms of thermal transport in polymers is of vital interest. Here, we leverage molecular layer deposition to grow three types of hybrid metalcone (i.e., alucone, zincone, and tincone) films and study their thermal and acoustic properties. The thermal conductivity of the hybrid polymer films ranges from 0.43 to 1.14 W m$^{-1}$ K$^{-1}$. Using kinetic theory, we trace the origin of thermal conductivity difference to sound speed change, which is dictated by the structural disorder within the films. Changing the disorder has negligible impacts on volumetric heat capacity and vibrational lifetimes. Our findings provide means to improve the thermal conductivity of organic, hybrid, and inorganic polymer films.

\end{abstract}



\section{Introduction}
The thermal, acoustic, and optical properties of polymers have been widely studied in literature due to their applications in electronic devices as flexible substrates, encapsulation layers, and insulating materials.\cite{duda2013thermal,chen2018directly,decoster2018density,li2019enhanced,nye2020understanding,ma2021pore} The thermal conductivity of polymers can span across three orders of magnitude; ranging from 0.2 W m$^{-1}$ K$^{-1}$ in PMMA\cite{xie2016thermal} to 23 W m$^{-1}$ K$^{-1}$ in Zylon HM.\cite{wang2013thermal} Structural disorder and type of interactions between the polymer chains generally control the thermal conductivity of polymers.\cite{singh2014high,kim2015high,xu2018thermal,lv2021effect,lv2021effect1} However, establishing a clear physical picture of thermal transport in polymers has been a challenge. Study of new types of polymers with different degree of structural disorder can shed light onto this aspect.

Only recently, molecular layer deposition (MLD)\cite{nye2020understanding} has been successful in growing different types of stable hybrid organic-inorganic metalcone polymers. MLD uses a bifunctional organic monomer and multifunctional inorganic precursor to grow the hybrid metalcone films.\cite{dameron2008molecular} In addition to the usual applications of polymers, hybrid polymers are also useful in organic light emitting diodes and lithium/sodium ion batteries.\cite{zhao2017inorganic,kaliyappan2020constructing,han2021water} Use of metalcone films in these devices has resulted in improved lifetimes and stability. However, due to the relatively new age of the metalcone films, their thermal and acoustic properties remain unexplored. 

In this study, we characterize the thermal and acoustic properties of three different types of metalcone films-alucone, zincone, and tincone-grown via MLD. The three metalcone films each possess a different degree of structural disorder (characterized by double reactions) based on growth. We use steady-state thermoreflectance (SSTR)\cite{braun2019steady,hoque2021thermal,hoque2021high} and time-domain thermoreflectance (TDTR)\cite{cahill2004analysis,feser2014pump,jiang2018tutorial} to measure the thermal conductivity and longitudinal sound speed of the films, respectively. Our measurements reveal that the structural disorder controls the longitudinal sound speed of the films. The sound speed, in turn, dictates the thermal conductivity of the metalcone films. We further verify the findings by measuring thermal conductivity, longitudinal sound speed, volumetric heat capacity, and lifetime of the vibration modes of an ordered and disordered alucone film. Our study opens up new pathways for tuning the thermal and acoustic properties of hybrid polymer films by manipulating disorder.

\section{Growth details of the metalcone films}

\begin{figure}[hbt!]
\centering
\includegraphics[scale = 0.45]{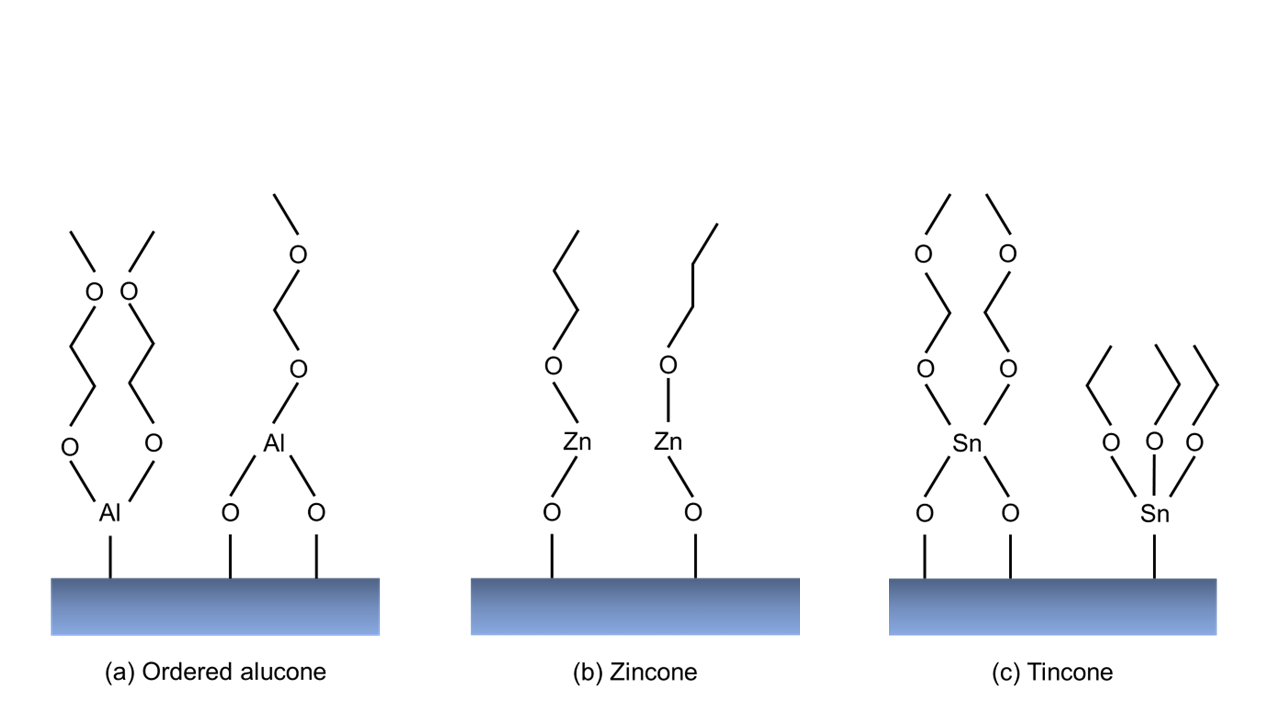}\\
\includegraphics[scale = 0.45]{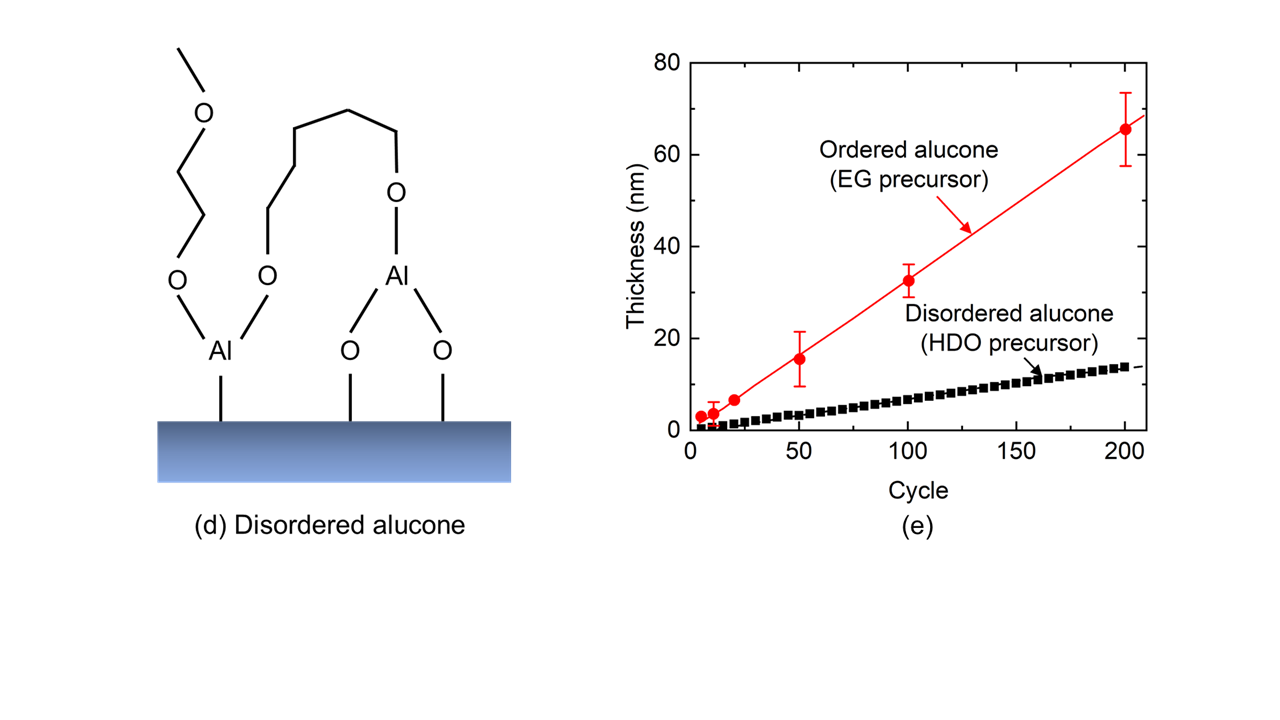}\\
\caption{Structures of the (a) ordered alucone, (b) zincone, (c) tincone, and (d) disordered alucone films. (e) Growth rate of the ordered and disordered alucone films measured by ellipsometry.}
\label{fig:1}
\end{figure}

The metalcone films are deposited via MLD using various metalorganic precursors and ethylene glycol (EG) as the coreactants. Alucone is deposited from trimethylaluminum (TMA) and EG. Zincone is deposited from diethyl zinc (DEZ) and EG.\cite{peng2009zincone} Tincone is deposited from tetrakis (dimethylamido) tin (TDMASn) and EG. Each precursor is handled only in nitrogen-ambient before being installed on the home-built cylindrical or spherical MLD reactors.\cite{nye2020understanding,nye2022situ} Precursors are heated as follows to achieve sufficient vapor pressure for deposition: TDMASn to 65 $^\circ$C, TMA to room temperature, DEZ to room temperature, and EG to 80 $^\circ$C for alucone and 70 $^\circ$C for tincone and zincone. The deposition is carried out at 100 $^\circ$C and 400 mTorr. The silicon (Si) substrates are cleaned in a piranha solution of 1:1 volume ratio H$_2$O$_2$:H$_2$SO$_4$ for 15 minutes before use. The structures of the metalcone films are shown in Figure 1. The thicknesses of the films range from 1 to 156 nm.

To study the effects of disorder on the thermal and acoustic properties of the metalcone films, we also deposit alucone films from TMA and 1,6-hexamethylenediol (HDO). While alucone films from TMA/EG is common, HDO represents a more unique and much less studied reactant. The alucone film grown from HDO precursors are less ordered compared to that of EG as discussed in the next section. As a result, we define the HDO alucone films as disordered in this study.

\section{Characterization of structural disorders in the films}

We characterize the disorders in the metalcone films by the phenomenon of double reactions (DRs).\cite{bergsman2017effect,nye2022situ} When DRs occur, both reactive groups of the homobifunctional precursors react with the growth surface during the same half-cycle without providing a new site for continued growth. DRs are more common for more flexible molecules (i.e., with longer aliphatic hydrocarbon chains). Among the metalcone films, tincone is expected to undergo a larger number of DRs during deposition compared to alucone because the tin precursor can bond with more organic ligands (4 and 3 for tincone and alucone, respectively). Increased DRs will lead to increased interactions between adjacent molecules/chains in a film, thereby increasing disorder in tincone compared to the alucone films.

Moreover, for the alucone films, the DRs are expected to be different based on the precursors (i.e., EG and HDO) used. EG is relatively short, and is expected to have few DRs and give rise to a relatively more ordered alucone film. On the other hand, the longer chain HDO precursor is expected to undergo more frequent DRs, thus resulting in more cross-linked or tangled chains during deposition and a more disordered film. The structure of a disordered alucone film is shown in Figure 1(d).

Figure 1(e) plots film thickness as a function of atomic layer deposition cycle for both alucone films (TMA/EG and TMA/HDO), as measured by spectroscopic ellipsometry. The growth rates correspond to 0.3 and 0.07 nm/cycle for alucone deposited using EG and HDO, respectively, consistent with previous results.\cite{george2011metalcones} The higher growth rate for the less flexible precursor (i.e. EG compared to HDO) is consistent with fewer expected double reactions for EG, as observed in many MLD processes.\cite{bergsman2017effect,bergsman2018mechanistic,nye2020understanding}

\section{Results and discussions}

\begin{figure}[b!]
\centering
\includegraphics[scale = 0.45]{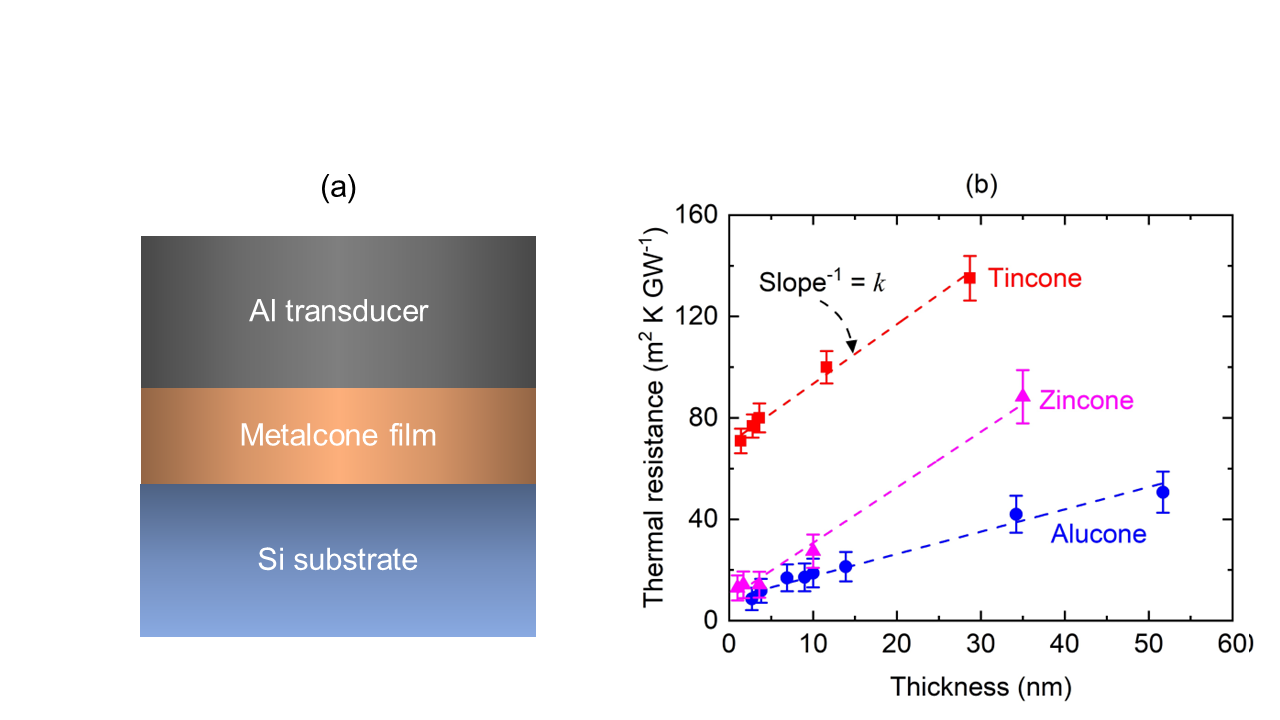}\\
\includegraphics[scale = 0.45]{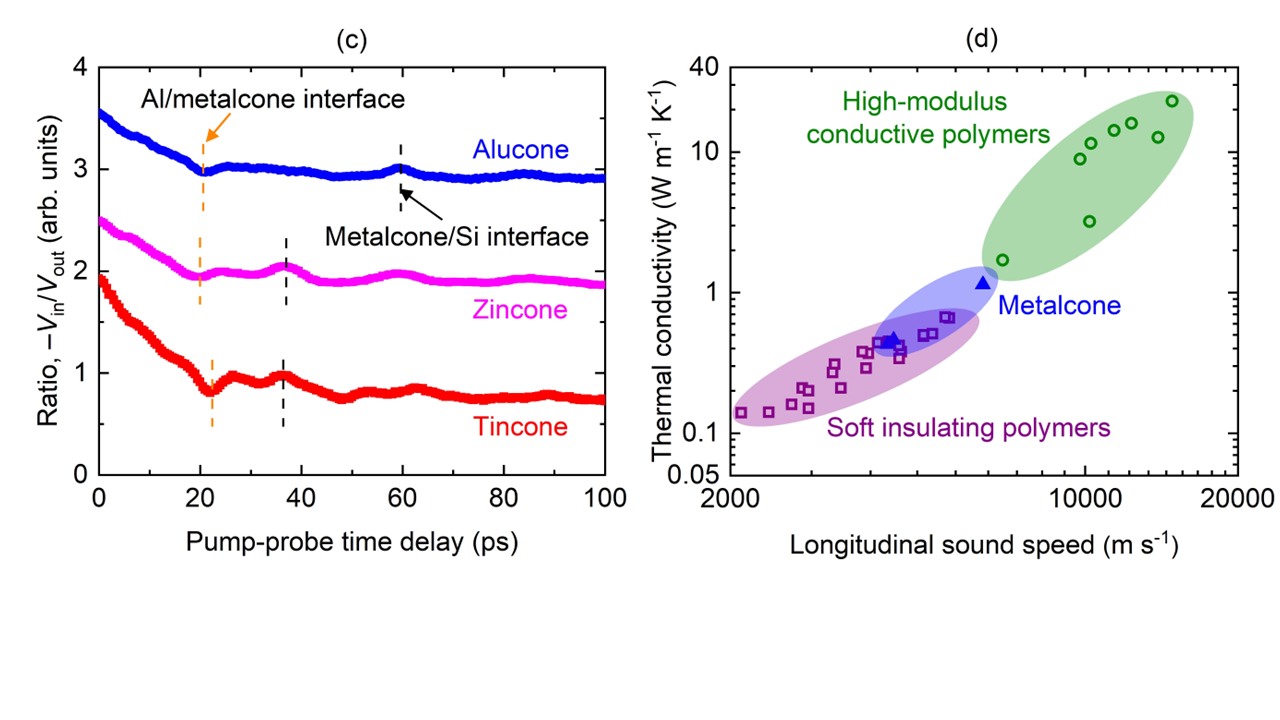}\\
\caption{(a) Schematic diagram of the sample geometry. (b) Thermal resistance as a function of alucone (EG precursor), zincone, and tincone film thickness. (c) Picosecond acoustic response of TDTR measurements for a 120, 35, and 29 nm alucone (EG precursor), zincone,  and tincone film, respectively. (d) Thermal conductivity vs longitudinal sound speed for a wide range of polymers. The data for the soft insulating polymers and high-modulus conductive polymers (hollow symbols) are taken from Refs\cite{wang2013thermal,xie2016thermal,xie2017high}.}
\label{fig:6_2}
\end{figure}

The thermal and acoustic properties of the metalcone films are characterized by laser-optical thermoreflectance techniques SSTR and TDTR, respectively. We use 1/e$^2$ pump and probe diameters of $\sim$20 $\mu$m for SSTR, whereas for TDTR, the diameters are $\sim$20 and 11 $\mu$m, respectively. Additional details of the techniques can be found in previous publications.\cite{jiang2018tutorial,braun2019steady,hoque2021high,hoque2021thermal,hoque2023interface} To convert the optical energy of the lasers into thermal energy, we deposit a thin aluminum film atop the samples via electron beam evaporation prior to the thermoreflectance measurements.\cite{decoster2019thermal} 

Figure 2(a) shows a schematic diagram of the sample geometry. As most of the metalcone films have thicknesses less than 50 nm, measuring thermal conductivity directly can be challenging. Therefore, we measure the total thermal resistance (\textit{R}) across the entire sample geometry via SSTR. The total thermal resistance can be expressed via the following series resistor model:\cite{decoster2019thermal}

\begin{equation}
\textit{R} = \bigg(\frac{1}{G}\bigg)_{Al/film} + \bigg(\frac{L}{\kappa}\bigg)_{film} + \bigg(\frac{1}{G}\bigg)_{film/Si}
\end{equation}

Where \textit{G}, $\kappa$, and \textit{L} represent thermal boundary conductance, thermal conductivity, and thickness of the metalcone films, respectively. Figure 2(b) shows the measured thermal resistances of the films as a function of thickness. As exhibited here, thermal resistance linearly  increases with film thickness. For such cases, thermal conductivity can be extracted from a linear fit to the thermal resistance as a function of film thickness. Here, the inverse of the slope ($\triangle$\textit{R}/$\triangle$\textit{L})$^{-1}$ provides the thermal conductivity of the films.\cite{aryana2020thermal} Using this methodology, we determine the thermal conductivity of the metalcone films as tabulated in Table 1.

\begin{table*}[hbt!]
\large
\centering
\renewcommand{\tablename}{\large Table}
\renewcommand{\thetable}{\large 1}
\renewcommand{\arraystretch}{1.2}
\caption{\large Thermal conductivity and longitudinal sound speed of the metalcone films.}
\begin{tabular}{cccc}
\hline
\hline
Metalcone films & Thermal conductivity & Longitudinal sound speed\\
& (W m$^{-1}$ K$^{-1}$) & (m s$^{-1}$)\\
\hline
Alucone (EG precursor) & 1.14 $\pm$ 0.18 & 6288 $\pm$ 324 \\
Zincone & 0.46 $\pm$ 0.06 & 4192 $\pm$ 401\\
Tincone & 0.43 $\pm$ 0.07 & 4080 $\pm$ 297 \\
\hline
\hline
\end{tabular}
 \end{table*}

We measure the longitudinal sound speed of the metalcone films via picosecond acoustics\cite{braun2016breaking,braun2018charge,gorham2014density,aryana2021interface,hoque2023interface,shankar2023microscale} using TDTR. Figure 2(c) shows the picosecond acoustic response of the metalcone films. As exhibited in Table 1, alucone films (EG precursor) have the highest sound speed, whereas tincone films have the lowest. We posit that the sound speed difference between the two metalcone films is stemming from the difference in disorders. As sound speed is an indicator of stiffness or elastic modulus of a material,\cite{braun2018charge} the disorder is also changing the stiffness of the metalcone films.      

According to the kinetic theory, $\kappa$ = $\frac{1}{3}$\textit{C}\textit{v}$^{2}\tau$, where \textit{C}, \textit{v}, and $\tau$ represent volumetric heat capacity, sound speed, and lifetime of the vibrational modes, respectively.\cite{braun2016breaking} Using this equation, the thermal conductivity difference among the metalcone films can be quantitatively explained by the difference in sound speed. This provides evidence that instead of volumetric heat capacity or lifetimes, sound speed is dictating the thermal conductivity of the metalcone films. A similar trend has been observed in literature\cite{wang2013thermal,xie2016thermal,xie2017high} for other polymers as illustrated in Figure 2(d). This figure also shows that the metalcone films can bridge an important gap between the soft, insulating polymers and high-modulus, conductive polymers. 

To further show the impact of disorders on the sound speed and thermal conductivity of the metalcone films, we measure $\kappa$, \textit{C}, \textit{v}, and $\tau$ of an ordered (EG precursor) and disordered (HDO precursor) alucone film. The TDTR multifrequency approach\cite{olson2018size,aryana2021suppressed} is used to simultaneously measure the thermal conductivity and volumetric heat capacity of 120 nm ordered and 156 nm disordered alucone films. As shown in Figure 3(a), the volumetric heat capacity remains the same for both cases: 2.0 $\pm$ 0.3 MJ m$^{-3}$ K$^{-1}$. The thermal conductivity, on the other hand, drops from 1.05 $\pm$ 0.15 W m$^{-1}$ K$^{-1}$ for the ordered film to 0.5 $\pm$ 0.08 W m$^{-1}$ K$^{-1}$ for the disordered film. The longitudinal sound speed also changes from 6288 $\pm$ 324 m s$^{-1}$ for the ordered film to 4532 $\pm$ 230 m s$^{-1}$ for the disordered film. The picosecond acoustic response of the disordered alucone film is shown in Supporting Figure S1. The different thicknesses of the ordered and disordered alucone films are not expected to have an impact on the thermal conductivity and sound speed measurements.\cite{nye2020understanding,gorham2014density}

\begin{figure}[hbt!]
\centering
\includegraphics[width=\textwidth]{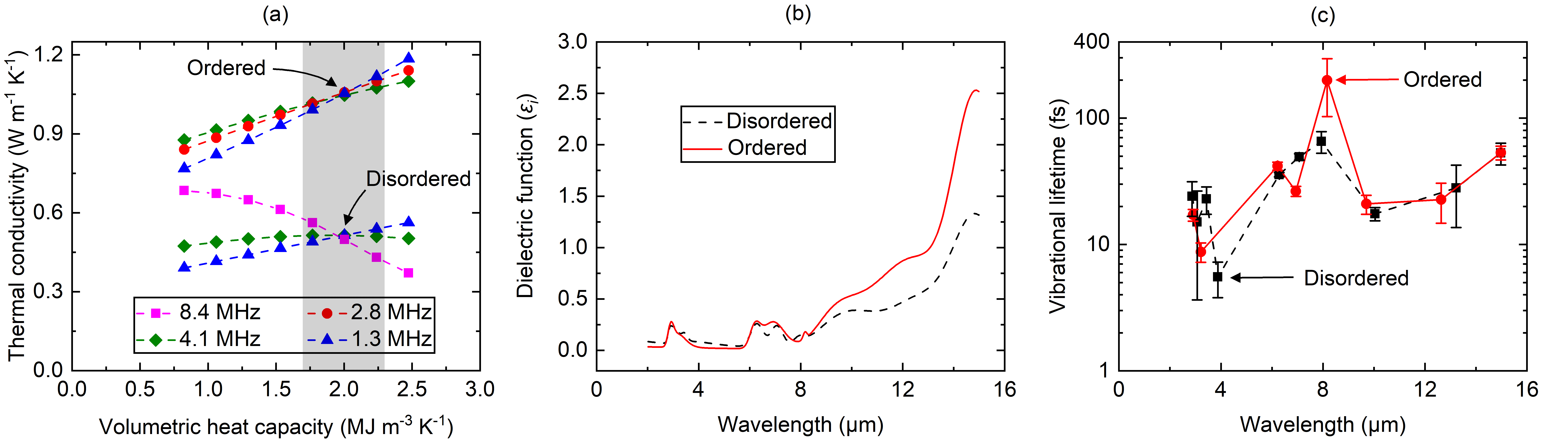}\\
\caption{(a) Thermal conductivity and volumetric heat capacity of the ordered and disordered alucone film. (b) Imaginary component of the dielectric function for the two cases as a function of wavelength. (c) Lifetime of the vibrational modes in alucone films. The uncertainty associated with the lifetime measurements are $\sim$10$\%$.}
\label{fig:6_2}
\end{figure}

The lifetimes associated with each vibrational mode in the alucone films are found from fitted oscillator models on ellipsometric data collected using IR-VASE (IR-VASE Mark II, J.A. Woollam Company). The ellipsometric data are acquired in the spectral range of 666-5000 cm$^{-1}$ (2-16 µm) with a resolution of 8 cm$^{-1}$. The raw ellipsometric data ($\Psi$ and $\Delta$) are processed using multiple Gaussian oscillators to determine the dielectric function of the alucone film. The parameters of each Gaussian oscillator, including amplitude, energy centroid, and broadening, are optimized to fit the model to the collected ellipsometric data (see Supporting Information). Figure 3(b) shows the imaginary component of the dielectric function as a function of wavelength for ordered and disordered alucone films. The optical lifetimes for the identified modes are determined by taking the reciprocal of the broadening parameter associated with each Gaussian oscillator. In Figure 3(c), we plot the derived lifetime of the vibrational modes for the two cases. As exhibited here, the lifetimes are nearly identical for ordered and disordered films at all wavelengths with one exception at 8.2 $\mu$m. The weighted average of lifetimes for the ordered film is 41 $\pm$ 4 fs, whereas for the disordered film, it is 38 $\pm$ 5 fs. This conclusively shows that the thermal conductivity difference between the two films originates from the sound speed difference, not volumetric heat capacity or lifetimes. Our results thus provide a clear picture of thermal transport in hybrid metalcone films. The thermal and acoustic properties reported here can be useful when incorporating metalcone films in electronic, photonic, and optoelectronic devices.   

\section{Conclusion}
In summary, we investigate the thermal and acoustic properties of alucone, zincone, and tincone films grown via MLD. Alucone possesses the highest thermal conductivity, while tincone the lowest. Structural disorder dictates the thermal transport of the hybrid metalcone films. By manipulating disorder and hence sound speed, thermal conductivity of the polymer films can be changed. Specifically, we find that films with a higher degree of disorder have a lower thermal conductivity. We further verify our conclusion by measuring the volumetric heat capacity and vibrational lifetime of a disordered and an ordered alucone film. The thermal transport mechanisms depicted in this study can also be applied to organic and inorganic polymer films.\\  

\textbf{Acknowledgement}\\

M.S.B.H, S.Z., and P.E.H. appreciate support from the Army Research Office, Grant Number W911NF-16-1-0406 and the National Science Foundation, Grant Number 2318576. R.A.N., S.A., S.W., and G.P. also acknowledge support form the Army Research Office, Grant Number W911NF-16-1-0406.\\

\newpage
\textbf{Supporting Information}\\

\textbf{S1. Metalcone deposition conditions}

During metalcone deposition, purified nitrogen (N$_2$ , 99.999$\%$, Arc3 Gases) is used as the carrier and purge gas. In the cylindrical reactor, tincone is deposited from TDMASn/EG following (10 s chamber evacuation/3 s TMDASn dose/60 s N$_2$ purge)/(0.3 s EG dose/60 s N$_2$ purge) and alucone is deposited from TMA/EG following (0.2 s TMA dose/60 s N$_2$ purge)/(0.2 s EG dose/60 s N$_2$ purge). In the spherical chamber, alucone is deposited from TMA/HDO following (0.4 s TMA dose/45 s N$_2$ purge)/(30 s chamber evacuation/8 s HDO dose/45 s N$_2$ purge).\\

\textbf{S2. Metalcone film characterizations}

Metalcone film thickness is measured with ellipsometry either in situ (Film Sense FS-1 multiwavelength ellipsometer) or ex situ (J. A. Woollam Co. alpha-SE spectroscopic ellipsometer) at an incidence angle of $\sim$70$^\circ$ relative to the surface normal. In situ data is collected at 436, 521, 599, and 638 nm while ex situ data is collected from 300 to 900 nm. Thickness is determined from a Cauchy model available with each ellipsometer’s software package. Uncertainty in film thickness is generally a few percent.\\

\textbf{S3. Picosecond acoustics of disordered alucone films}

\begin{figure}[hbt!]
\centering
\renewcommand{\thefigure}{S1}
\includegraphics[scale = 0.39]{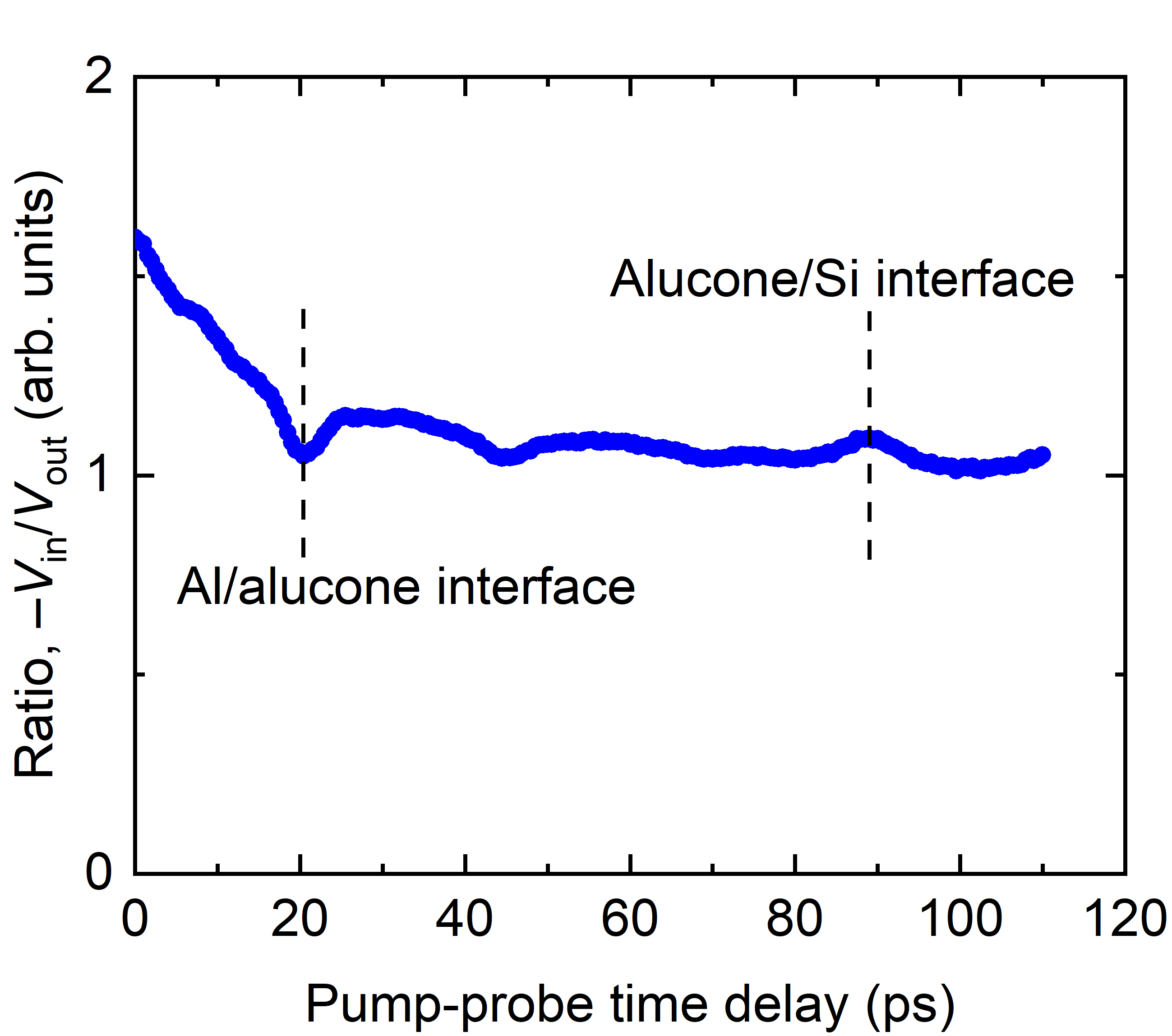}\\
\caption{Picosecond acoustic response of the 156 nm disordered alucone film.}
\label{fig:S1}
\end{figure}

\newpage
\begin{table*}[hbt!]
\large
\centering
\renewcommand{\tablename}{\large Table}
\renewcommand{\thetable}{\large S1}
\renewcommand{\arraystretch}{1.2}
\caption{\large Thermal conductivity and longitudinal sound speed of the polymers shown in Figure 2(d).}
\begin{tabular}{ccc}
\hline
\hline
Polymers & Thermal conductivity & Longitudinal sound speed\\
  &  (W m$^{-1}$ K$^{-1}$) & (m s$^{-1}$)\\
\hline
PVA & 0.31 & 3210\cite{xie2016thermal} \\
PAA & 0.37 & 3740\cite{xie2016thermal} \\
PVP & 0.27 & 3180\cite{xie2016thermal} \\
PAM & 0.38 & 4340\cite{xie2016thermal} \\
PSS & 0.38 & 3640\cite{xie2016thermal} \\
MC & 0.21 & 2770\cite{xie2016thermal} \\
PMMA & 0.20 & 2850\cite{xie2016thermal} \\
PAP & 0.16 & 2640\cite{xie2016thermal} \\
PALi & 0.55 & 5100\cite{xie2017high} \\
PANa & 0.45 & 4100\cite{xie2017high} \\
PACa & 0.49 & 4800\cite{xie2017high} \\
PAFe & 0.51 & 5000\cite{xie2017high} \\
PACu & 0.5 & 4800\cite{xie2017high} \\
PVPA & 0.44 & 3900\cite{xie2017high} \\
PVPLi & 0.63 & 5300\cite{xie2017high} \\
PVPMg & 0.66 & 5400\cite{xie2017high} \\
PVPCa & 0.67 & 5300\cite{xie2017high} \\
PVSNa & 0.42 & 4300\cite{xie2017high} \\
PDDA & 0.29 & 3700\cite{xie2017high} \\
PAH & 0.34 & 4300\cite{xie2017high} \\
PMPC & 0.21 & 3300\cite{xie2017high} \\
PS & 0.141 & 2380\cite{xie2017high} \\
ADP & 0.15 & 2850\cite{xie2017high} \\
DSQ & 0.14 & 2100\cite{xie2017high} \\
Vectra & 1.7 & 6879\cite{wang2013thermal} \\
Kevlar & 3.2 & 10203\cite{wang2013thermal} \\
M5 AS & 8.9 & 9772\cite{wang2013thermal} \\
Spectra 900 & 11.5 & 10275\cite{wang2013thermal} \\
Spectra 2000 & 16 & 12343\cite{wang2013thermal} \\
Dyneema & 14.2 & 11404\cite{wang2013thermal} \\
PBT & 12.7 & 13929\cite{wang2013thermal} \\
Zylon AS & 18.5 & 12187\cite{wang2013thermal} \\
Zylon HM & 23 & 14850\cite{wang2013thermal} \\
\hline
\hline
\end{tabular}
 \end{table*}

\newpage
\begin{table*}[hbt!]
\large
\centering
\renewcommand{\tablename}{\large Table}
\renewcommand{\thetable}{\large S2}
\renewcommand{\arraystretch}{1.2}
\caption{\large Gaussian oscillator parameters used to fit the ordered alucone film data, resulting in a mean square error of 1.36.}
\begin{tabular}{cccccc}
\hline
\hline
Oscillator No. & Amplitude & Centroid Frequency & Centroid Wavelength & Broadening & Lifetime\\
  &  & (cm$^{-1}$) & (um) & (cm$^{-1}$) & (fs)\\
\hline
1 & 0.19 & 3439.7 & 2.91 & 310.0 & 17.13 $\pm$ 1.82\\
2 & 0.13 & 3110.1 & 3.22 & 606.6 & 8.75 $\pm$ 1.52\\
3 & 0.22 & 1607.6 & 6.22 & 127.2 & 41.75 $\pm$ 3.04\\
4 & 0.26 & 1441.1 & 6.94 & 201.0 & 26.42 $\pm$ 2.43\\
5 & 0.07 & 1224.6 & 8.17 & 26.7 & 198.81 $\pm$ 96.13\\
6 & 0.45 & 1030.8 & 9.70 & 253.8 & 20.92 $\pm$ 3.61\\
7 & 0.85 & 791.65 & 12.63 & 234.4 & 22.65 $\pm$ 7.93\\
8 & 2.12 & 667.83 & 14.97 & 99.6 & 53.31 $\pm$ 6.74\\
\hline
\hline
\end{tabular}
 \end{table*}
\vspace{15mm}
\begin{table*}[hbt!]
\large
\centering
\renewcommand{\tablename}{\large Table}
\renewcommand{\thetable}{\large S3}
\renewcommand{\arraystretch}{1.2}
\caption{\large Gaussian oscillator parameters used to fit the disordered Alucone film, resulting in a mean square error of 0.37.}
\begin{tabular}{cccccc}
\hline
\hline
Oscillator No. & Amplitude & Centroid Frequency & Centroid Wavelength & Broadening & Lifetime\\
  &  & (cm$^{-1}$) & (um) & (cm$^{-1}$) & (fs)\\
\hline
1 & 0.11 & 3487.6 & 2.87 & 220.9 & 24.03 $\pm$ 7.32\\
2 & 0.15 & 3259.1 & 3.07 & 352.8 & 15.05 $\pm$ 11.41\\
3 & 0.08 & 2912.4 & 3.43 & 231.6 & 22.92 $\pm$ 5.57\\
4 & 0.04 & 2578.4 & 3.88 & 960.3 & 5.53 $\pm$ 1.73\\
5 & 0.23 & 1590.7 & 6.29 & 148.9 & 35.66 $\pm$ 1.38\\
6 & 0.21 & 1414.2 & 7.07 & 107.8 & 49.26 $\pm$ 2.83\\
7 & 0.07 & 1259.6 & 7.94 & 81.1 & 65.43 $\pm$ 12.79\\
8 & 0.36 & 995.6 & 10.04 & 303.6 & 17.48 $\pm$ 2.1\\
9 & 0.46 & 756.3 & 13.22 & 189.6 & 28.01 $\pm$ 14.41\\
10 & 1.04 & 667.8 & 14.98 & 100.2 & 52.99 $\pm$ 10.37\\
\hline
\hline
\end{tabular}
 \end{table*}

\medskip

%
\newpage
\bibliographystyle{achemso}
\bibliography{References}




\end{document}